%% file: main_usenix.tex
\documentclass[letterpaper,twocolumn,10pt]{article}
\usepackage{usenix}

\usepackage{tikz}
\usepackage{amsmath}
\usepackage{xspace}
\usepackage[nohyperlinks,nolist]{acronym}
\usepackage{booktabs}
\usepackage{todonotes}
\usepackage{siunitx}
\usepackage{array}
\usepackage{ragged2e}
\usepackage{tabularx}  
\newcolumntype{Y}{>{\RaggedRight\arraybackslash}X} 
\newcolumntype{C}{>{\centering\arraybackslash}X}
\newcolumntype{N}{>{\centering\arraybackslash}>{\hsize=.5\hsize}X}
\usepackage{tikz}
\def\checkmark{\tikz\fill[scale=0.4](0,.35) -- (.25,0) -- (1,.7) -- (.25,.15) -- cycle;} 

\newcommand{\name}{5Gone\xspace}

\begin{document}

\date{}

\title{\Large \bf 5Gone: Uplink Overshadowing Attacks in 5G-SA}

\author{
{\rm Simon Erni}\\
ETH Zurich
\and
{\rm Martin Kotuliak}\\
ETH Zurich
\and
{\rm Marc Roeschlin}\\
Unaffiliated
\and
{\rm Richard Baker}\\
University of Oxford
\and
{\rm Srdjan Capkun}\\
ETH Zurich
} %

\maketitle
\input{acronyms}

\renewcommand{\sectionautorefname}{Section}
\renewcommand{\subsectionautorefname}{Section}
\renewcommand{\subsubsectionautorefname}{Section}

\input{content/00_abstract}

\input{content/01_introduction}

\input{content/02_background}

\input{content/03_threat_challenges}

\input{content/05_attack} %
\input{content/04_implementation}

\input{content/06_evaluation}
\input{content/07_related_work}

\input{content/08_countermeasures} %
\input{content/09_conclusion} %

\cleardoublepage
\appendix
\input{content/a0_ethics}
\cleardoublepage

\input{content/a1_openscience}
\cleardoublepage
\bibliographystyle{plain}
\bibliography{simon,dirac,richard,relwork}
\cleardoublepage

\end{document}

%% file: acronyms.tex
    \newacro{5G}{Fifth-Generation Cellular}
    \newacro{AMF}{Access and Mobility Management Function}
    \newacro{FBS}{Fake Base Station}
    \newacro{DoS}{Denial-of-Service}
    \newacro{5G-NSA}{5G Non-Standalone}
    \newacro{5G-SA}{5G Standalone}
    \newacro{NB-IoT}{Narrowband-Internet of Things}
    \newacro{UE}{User Equipment}
    \newacro{CE}{Control Element}
    \newacro{SDR}{Software Defined Radio}
    \newacro{GPSDO}{GPS Disciplined Oscillator}
    \newacro{PRACH}{Physical Random Access Channel}
    \newacro{RAN}{Radio Access Network}

    \newacro{CCCH}{Common Control Channel}
    \newacro{DCCH}{Dedicated Control Channel}
    \newacro{PLMN}{Public Land Mobile Network}
    \newacro{PDCP}{Packet Data Convergence Protocol}
    \newacro{RLC}{Radio Link Control}
    \newacro{SDU}{Service Data Unit}
    \newacro{gNB}{Next Generation Node B}
    \newacro{5GC}{5G Core Network}
    \newacro{URLLC}{Ultra Reliable and Low Latency Communication}
    \newacro{RA}{Random Access}
    \newacro{RAR}{Random Access Response}
    \newacro{RRC}{Radio Resource Control}
    \newacro{PHY}{Physical}
    \newacro{NAS}{Non-Access Stratum}
    \newacro{NGAP}{NG Application Protocol}
    \newacro{MAC}{Medium Access Control}
    \newacro{MIB}{Master Information Block}
    \newacro{SIB}{System Information Block}
    \newacro{DCI}{Downlink Control Information}
    \newacro{RNTI}{Radio Network Temporary Identifier}

    \newacro{TMSI}{Temporary Mobile Subscriber Identity}
    \newacro{PUSCH}{Physical Uplink Shared Channel}
    \newacro{PDCCH}{Physical Downlink Control Channel}
    \newacro{PDSCH}{Physical Downlink Shared Channel}
    \newacro{SUCI}{Subscription Concealed Identifier}
    \newacro{SUPI}{Subscription Permanent Identifier}

    \newacro{IRB}{Institutional Review Board}

%% file: content/00_abstract.tex
\global\csname @topnum\endcsname 0
\global\csname @botnum\endcsname 0

\begin{abstract}
5G presents numerous advantages compared to previous generations: improved throughput, lower latency, and improved privacy protection for subscribers. Attacks against 5G standalone (SA) commonly use fake base stations (FBS), which need to operate at a very high output power level to lure victim phones to connect to them and are thus highly detectable. In this paper, we introduce 5Gone, a powerful software-defined radio (SDR)-based uplink overshadowing attack method against 5G-SA. 5Gone exploits deficiencies in the 3GPP standard to perform surgical, covert denial-of-service, privacy, and downgrade attacks. Uplink overshadowing means that an attacker is transmitting at exactly the same time and frequency as the victim UE, but with a slightly higher output power. 5Gone runs on a COTS x86 computer without any need for dedicated hardware acceleration and can overshadow commercial 100 MHz cells with an E2E latency of less than 500$\mu$s, which up to now has not been possible with any software-based UE implementation. We demonstrate that 5Gone is highly scalable, even when many UEs are connecting in parallel, and finally evaluate the attacks end-to-end against 7 phone models and three different chipset vendors both in our lab and in the real-world on public gNodeBs.

\end{abstract}

%% file: content/01_introduction.tex
\begin{figure}[t]
	\centering
	\includegraphics[width=\linewidth]{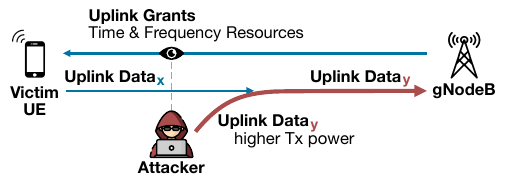}
	\caption{Illustration of an uplink overshadowing attack. An attacker transmits on the same uplink time and frequency resources as the victim UE, but with a slightly higher transmit power. The gNB will only decode the stronger signal, thus the attacker can change the data transmitted by the UE without the UE knowing anything about it.}
	\label{fig:overview}
\end{figure}

\section{Introduction}

5G has brought with it an array of improvements to make it applicable to new use-cases, such as high-bandwidth applications, large-scale IoT, and low-latency, high-reliability device applications. Although cellular networks have been recognized as critical infrastructure for decades~\cite{EO13010_1996}, with so many vitally-important cellular systems depending on security and availability, the need for close scrutiny is even clearer.

A number of attacks impacting availability and subscriber privacy have been presented against \acs{5G-SA}. These attack methods include smart- or naïve- jamming, fake base-stations~\cite{chlosta_5g_2021} or overshadowing of the downlink spectrum~\cite{ludant_sigunder_2021, luo_sni5gect_2025, hu_sigoverlay_2025}. However, all of these attacks are competing with the usually very high output power of a base station. It has been shown in prior work -- focused on 4G LTE or NB-IoT -- that uplink overshadowing is a highly effective attack method~\cite{tan_breaking_2022, erni_adaptover_2022, erni_glados_2025}, since uplink overshadowing attacks only require transmission power greater than \textit{that of a phone}. As such, these previous works demonstrate that it is possible for an attacker nearly 4 km away to launch targeted attacks on a UE that is right next to a base station. Furthermore, overshadowing attacks surgically replace single messages sent over the radio layer, making them significantly harder to detect compared to classical methods. Decreased power requirements, increased range, and stealthiness make uplink overshadowing a substantially more severe threat compared to conventional cellular attack vectors.

In this work, we show that an uplink overshadowing attacker also constitutes a practical and significant threat to 5G standalone (5G-SA) networks. Such an attacker can achieve an impact comparable to that of a \ac{FBS}, while avoiding the well-known limitations associated with such \ac{FBS} attacks. We design, implement, and evaluate three classes of uplink overshadowing attacks: (i) selective and non-selective cell-wide denial-of-service and downgrade attacks, (ii) a privacy-violating attack that enables the extraction of the \ac{SUCI}, and (iii) a SUCI replay attack that is capable of linking captured SUCIs to current UE connections, or capable of launching a denial-of-service/downgrade attack while still allowing the attacker to communicate.

Attacking \acs{5G-SA} via uplink overshadowing presents greater complexity compared to 4G, mainly due to factors such as a maximum cell bandwidth of 100 MHz (compared to 20 MHz for LTE) and extremely low-latency response requirements for devices or attackers of less than 500$\mu$s (compared to 4ms for LTE). None of the open source software UE implementations previously used for this class of attackers are capable of handling cells with these challenging configurations. Yet, from our observations, these are the configurations typically seen in real deployments.

For this purpose, we introduce \name, an uplink overshadowing attack platform for 5G-SA cellular networks. \name solves these challenges and demonstrates attacking 64 UEs in parallel on a 100 MHz cell, while reacting with correctly encoded uplink transmissions in as little as 343$\mu$s. These results show that \name is fast enough for even the most aggressive latency configurations and dense urban loads seen in the real deployments.

To summarize, in this paper, we make the following contributions:
\begin{itemize}
    \item We are the first to present and validate an uplink overshadowing attack method against 5G-SA, capable of targeted or cell-wide denial-of-service, as well as downgrade and privacy attacks.
    \item We introduce \name, the first uplink overshadowing attack platform capable of operating in real-world 5G-SA deployments, overcoming key challenges such as 100 MHz cell bandwidths and sub-500$\mu$s reaction time requirements. We show its real-time performance in extensive measurements and describe its architecture.
    \item We evaluate the attacks in both lab and real-world conditions with the permission of the operator against our own phones, demonstrating the practicality and effectiveness of the uplink overshadowing attacks, and we publish detailed attack logs and traces from all involved components.
    
\end{itemize}

%% file: content/02_background.tex
\section{Background}
In 5G, users connect to the base stations, also called \ac{gNB} or gNodeB, using a \ac{UE}. The \ac{gNB} provides the \ac{RAN} portion of the service and connects onwards to the network core, called a \ac{5GC}, which includes a range of services (e.g., subscriber management, roaming, billing) and handles onward traffic to public telephone networks and the Internet. 

\begin{figure}
	\centering
	\includegraphics[width=\linewidth]{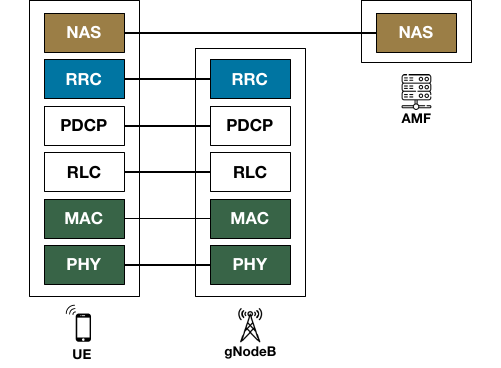}
	\caption{Elements of the 5G protocol stack that are relevant to uplink overshadowing attacks presented in this paper.}
	\label{fig:protocol_stack}
\end{figure}

\subsection{Connection Establishment}\label{sec:contention_resolution}
\begin{figure}
	\centering
	\includegraphics[width=\linewidth]{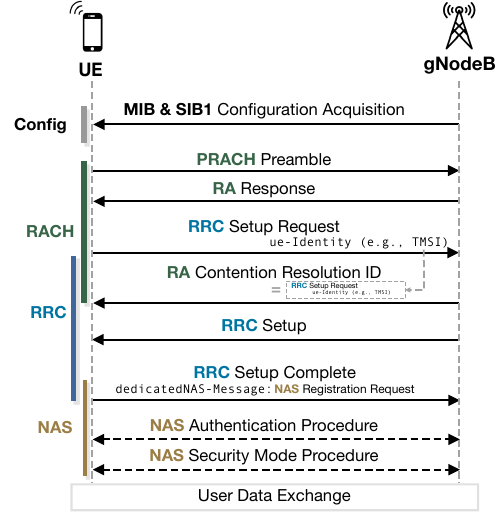}
	\caption{Sequence Diagram of a UE connecting to the network for the first time, e.g., after rebooting or toggling flightmode.}
	\label{fig:background_registration}
\end{figure}

In~\autoref{fig:background_registration}, a connection establishment procedure is shown for a \ac{UE} that has just been turned on.

First, the UE scans for available 5G cells in the spectrum. It does so by decoding the broadcasted configurations in the \ac{MIB} and \ac{SIB} messages. These messages contain information on which operator is running this cell, as well as other relevant configuration parameters necessary to connect to this cell.

Once the UE has decided to proceed with registration on a cell, it starts a random access procedure. It first sends a \aclu{RA} Preamble on the time and frequency resources configured in the \ac{SIB}. Once the \ac{gNB} has received such a preamble, it responds with a \ac{RAR} message, containing the preamble index and assigning a random 16-bit \ac{RNTI} to the UE. From here onward, all communication on the downlink is linked to this \ac{RNTI}. The RAR also contains an uplink grant, allowing the UE to send the next uplink message (Msg3). The UE uses this grant to transmit the \texttt{\ac{RRC} Setup Request}, containing any previous identifier the UE may have from an earlier connection (e.g., TMSI), or a random 39-bit number if none exists.

There is only a limited set of \ac{RA} preamble indices available for a \ac{UE}. For this reason, 5G includes a \textit{contention resolution} mechanism that reflects the \texttt{RRC Setup Request} message from the UE on the downlink in the form of the Contention Resolution ID. If there is more than one UE trying to connect with the same preamble, only one UE will receive a Contention Resolution ID that matches its RRC Setup Request. For all others, the contention resolution ID will not match, signaling to them that there has been a \textit{contention} and that they should drop the connection attempt and try again.

Once the contention resolution process is complete, the \ac{gNB} transmits further configuration information to the \ac{UE} using the RRC Setup message. The UE confirms the successful setup by applying the configuration and responding with a \texttt{RRC Setup Complete} message. Embedded within is a first higher layer message of the \ac{NAS} layer. \ac{NAS} messages are actually relayed by the gNB to the core.

To establish a connection with the core, the UE transmits its previously assigned temporary identifier (\ac{TMSI}) or an encrypted persistent identifier (\ac{SUCI}) in the \texttt{NAS Registration Request} message. If the core can find the supplied identifier in its database, the establishment continues with an authentication procedure, during which the encryption keys for this session are established. This process finishes with the security mode procedure, in which encryption and integrity protection algorithms are activated and used henceforth. Finally, user traffic (voice, data) can be exchanged.

\subsection{Overshadowing Attacks}

The approach of overshadowing in cellular networks was introduced in~\cite{yang_hiding_2019} and further refined into uplink overshadowing in~\cite{erni_protocol-aware_2020,tan_breaking_2022}. However, up to now, no implementation has been shown to be feasible in 5G-SA networks.

\name introduces uplink overshadowing in 5G-SA networks. The basic concept of the attack method is the same as in previous works and operates as follows. An attacker located in the cell coverage area collects all control messages destined for all UEs connecting to the cell. The control messages also include uplink allocations, indicating where the UEs are allowed and expected to transmit. The attacker then sends a correctly encoded and modulated message on the same time and frequency resources, but with a slightly higher output power. At the base station, only the stronger signal is decoded; hence, the attacker is able to replace messages from the UE over-the-air.

%% file: content/03_threat_challenges.tex
\section{Assumptions and Challenges}

We consider an attacker with commodity \ac{SDR} equipment, desktop-grade computers and a suitable software stack for processing radio samples in accordance with the 5G/NR standards. This is possible using existing open libraries and the modifications described in this paper. The attacker has no access to or knowledge of any cryptographic material belonging either to the victim or the network (provider). 

The attacker is capable of receiving the downlink of a 5G base station to which the victim is trying to attach. This reception must be of sufficient quality to decode the initial control messages that are sent to the victim. Furthermore, we require that the attacker is capable of transmitting uplink messages with enough power such that the uplink messages from a victim UE are overshadowed at the \ac{gNB}'s antenna. This implies that when an attacker is far away from a base station, it requires more transmit power than when being right next to a base station. The attacker is also assumed to be capable, when necessary, of listening on the uplink and again with sufficient quality to decode early control messages.

Several key challenges are specific to 5G-SA and need to be solved by any 5G-SA uplink overshadowing attacker. First, the bandwidth of 5G-SA has increased five-fold from a maximum of 20 MHz in LTE to 100 MHz. This requires hardware and software with sufficient performance to receive and transmit at suitable sampling rates. Furthermore, in production 5G-SA networks, if a UE receives an uplink grant in slot $x$, it might need to transmit in the slot $x+1$. Since the slot duration is 500$\mu$s (for 30 kHz subcarrier-spacing), this leaves less than 500$\mu$s for the UE to perform \ac{DCI} decoding, uplink payload generation, \ac{PUSCH} coding, modulation, and transmission to the radio. We compare our work to open source software UE implementations in~\autoref{sec:related_work} and find that none of the existing software is capable of achieving this low-latency up to now.

%% file: content/05_attack.tex
\section{5G Uplink Overshadowing Attacks}\label{sec:attacks}

In this section, we showcase attacks on 5G SA that are made possible using uplink overshadowing. 
We first describe an attack capable of entirely disabling all connections to a 5G base station in a stealthy manner. Next, we describe a downgrade attack that coerces UEs to fall back to earlier, less secure access technologies, followed by a privacy attack that extracts the \ac{SUCI} from connecting UEs and links it to their \ac{TMSI}, and finally, a \ac{SUCI} replay attack that can test if a \ac{SUCI} belongs to a subscriber and can cause selective denial-of-service/downgrade attack that prevents the phone from connecting to any cellular network, regardless of technology.

\subsection{Cell-Wide Denial-of-Service Attack}\label{sec:mac_dos}

\begin{figure}
	\centering
	\includegraphics[width=\linewidth]{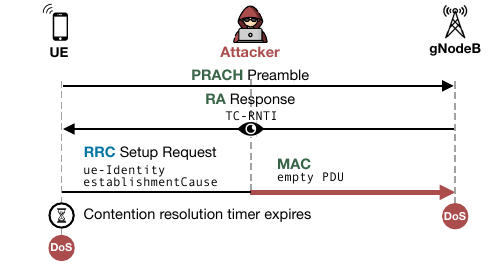}
	\caption{Attack on the Random-Access Procedure causing a Downgrade by preventing any connection establishment to a gNodeB.}
	\label{fig:dos_mac}
\end{figure}

This attack, shown in~\autoref{fig:dos_mac}, intercepts all connection establishments at the earliest possible moment and causes all random access procedures between the \ac{UE} and \ac{gNB} to fail, resulting in a cell-wide denial-of-service.

The attack works by first listening on the downlink for an \texttt{RA Response} message, which announces the presence of a UE connecting to the cell. The \texttt{RA Response} contains the \texttt{TC-RNTI}, as well as an uplink grant for the UE to transport its \textit{Msg3}. The attacker uses the uplink grant from the \texttt{RA Response} to transmit its own message to the gNodeB with slightly higher output power, overshadowing the original \textit{Msg3} from the UE.

This overshadowing of the \textit{Msg3} might happen in legitimate scenarios; for example, if two UEs connect at the same time and choose the same PRACH preamble, they will both react to the same \texttt{RA Response}, with one UE overshadowing the other and causing a \textit{contention}. Only after receiving the \texttt{Contention Resolution ID}, which will be a copy of the transported \textit{Msg3}, will the UE consider the random access procedure to have succeeded and will continue with the connection establishment. The other UE whose \textit{Msg3} was overshadowed will immediately abort the connection establishment and re-try.

In this attack, we are exploiting this contention resolution mechanism to block and terminate all connections before any control or user data is transmitted. As defined by 3GPP~\cite{3gpp_ts_38321_3gpp_2025}, the Msg3 contains a C-RNTI \ac{MAC} \ac{CE} or a \ac{CCCH} \ac{SDU}. In our attack, we are sending neither of these and are just transmitting an empty MAC PDU containing nothing but padding bytes. In our observations of various \ac{gNB}s from both operators and our Amarisoft, the \ac{gNB} drops the connection attempt and does not send any further uplink grants or messages. This will cause the contention resolution timer of the UE to expire, which, according to 3GPP~\cite{3gpp_ts_38321_3gpp_2025}, results in a failed Contention Resolution procedure and a re-transmission of the Random Access Preamble. After transmitting the Preamble at most 200 times (in practice much lower, based on the cell configuration), it will restart the Random Access Resource selection procedure, potentially after a backoff of at most 1.92 seconds. However, eventually it will stop attempting to connect to the cell and select a different cell~\cite{3gpp_ts_38331_3gpp_2025,3gpp_ts_38304_3gpp_2025}. %

In case the base station does not stop the connection attempt after receiving an empty MAC PDU, alternatively, the attacker could attempt to send an invalid MAC packet, triggering a parsing error, or adhere to the specification~\cite{3gpp_ts_38321_3gpp_2025} by transmitting a valid \ac{RRC}-\ac{CCCH} (e.g., an \texttt{RRC Connection Request}) or a C-RNTI \ac{MAC} \ac{CE}, both of which will cause the \ac{gNB} to transmit an invalid contention resolution ID to the victim UE. %

The benefit of this attack is that it is both highly impactful and stealthy and, in comparison with higher-layer attacks, executable with little resources required from the attacker. First, it is impactful because it successfully denies any and all connections of UEs attempting to connect to a \ac{gNB}, regardless of whether it is due to a handover or a fresh connection establishment to the cell. Additionally, it is stealthy since it is purely constrained to the \ac{PHY} and \ac{MAC} layers. This means that no messages over the \ac{NGAP} interface are exchanged since no \ac{RRC} or even \ac{NAS} message is exchanged at all. Also, since nothing on the downlink is modified, the presence of such an attacker cannot be easily determined by a bystander, who would easily be able to spot, e.g., fake base station attacks by analyzing broadcast information. Finally, since the attacker only needs to decode the random access response broadcast channel, it requires only minimal resources. 
We evaluated the attack in~\autoref{sec:eval_mac_dos}, where we show how \name can attack more than 400 connection attempts per second on a 100 MHz cell using only one CPU core for the cell-level processing.

A use case for such an attack presents itself in the form of a defense against \ac{FBS}. Once an \ac{FBS} is detected, this cell-wide DoS attack could be launched against it to prevent all UEs in the vicinity from connecting to it, effectively \textit{protecting} their privacy and availability.

\subsection{Registration Reject Downgrade Attack}\label{sec:registration_reject}

In this attack, we show how receiving a Registration Reject in NAS results in a downgrade attack, coercing the phone to connect to older, less secure technologies.

\begin{figure}[h]
	\centering
	\includegraphics[width=\linewidth]{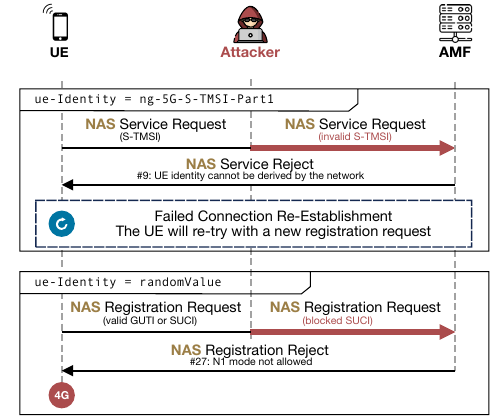}
	\caption{Registration Reject Downgrade Attack. In the upper half, the UE has a pre-existing session with the AMF and would like to resume it, which is blocked with a Service Reject attack and clears the session in the UE. In the botom half, the UE attempts to register to the network, but is immediately blocked with the cause value \textit{N1 mode not allowed}, prompting the UE to downgrade to 4G.}
	\label{fig:registration_downgrade}
\end{figure}

In~\autoref{fig:registration_downgrade}, we show the attack procedure, starting with the UE having a pre-existing session with the \ac{AMF} and ending with the UE downgrading to 4G. First, the attacker overshadows the \texttt{NAS Service Request} of the UE, replacing it with the same message type but containing an invalid S-\ac{TMSI}. Since this S-TMSI cannot be mapped to an existing session, the \ac{AMF} aborts the connection establishment with a \texttt{NAS Service Reject} message and the UE immediately retries registration.

When now re-establishing an \ac{RRC} connection, the UE will send an \texttt{RRC setup request} containing a randomValue instead of the S-TMSI. Since the \texttt{RRC setup request} is reflected on the downlink as the contention resolution ID, the attacker can anticipate that the UE will send a \texttt{NAS Registration Request} in the \texttt{RRC setup complete} message. Thus, the attacker overshadows it and place a \ac{SUCI} inside the \texttt{NAS Registration Request} that is not allowed to receive 5G-SA services in this cell. This SUCI can be acquired by, e.g., obtaining a subscription without any 5G-SA support, obtaining an expired SIM card, or by generating random IMSIs. In either case, the chosen SUCI is not allowed to obtain 5G services, and the AMF replies with the desired \texttt{NAS Registration Reject}. It depends entirely on the implementation of the AMF which reject code is used; in our experiments with the operator, this was often cause value \#27, meaning \textit{N1 mode not allowed}, or \# 15, meaning \textit{No suitable cells in tracking area}, forbidding the UE to connect to any cells in the tracking area. As specified by 3GPP~\cite{3gpp_ts_24501_3gpp_2025}, for these codes the UE will consider itself blocked across the entire tracking area (a large collection of cells belonging to the same geographical area) or across the entire 5G network, both of which ultimately cause a downgrade to LTE.

\subsection{SUCI Extraction Attack}\label{sec:suci_extraction}

In 5G, the persistent subscriber identity IMSI, or \ac{SUPI}, can be transmitted in an encrypted form, called the \ac{SUCI}. Still, previous works~\cite{basin_formal_2018,chlosta_5g_2021} have shown that, after obtaining a SUCI, it is possible to test whether a connecting UE belongs to this SUCI, even weeks or months later. However, during normal operations, the \ac{SUCI} is only transmitted in a very limited manner, e.g., after turning on a phone. Nevertheless, 3GPP standards~\cite{3gpp_ts_24501_3gpp_2025} still allow the SUCI to be requested at any moment using a \texttt{NAS Identity Request} message, which we exploit in this attack. Note that this is the only attack where, in addition to the downlink receiver and uplink transmitter, we also require an uplink sniffer.

\begin{figure}[h]
	\centering
	\includegraphics[width=\linewidth]{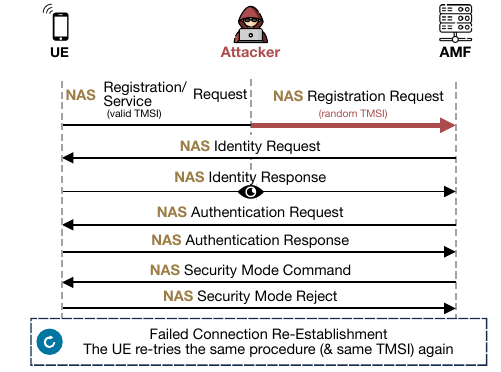}
	\caption{SUCI Extraction Attack by triggering an \texttt{NAS Identity Request} message.}
	\label{fig:suci_phase_1}
\end{figure}

In this attack, it does not matter whether the UE connects with a \texttt{Service} or \texttt{Registration} request, in either case, the attacker replaces the uplink NAS message with a \texttt{NAS Registration Request} that contains a random invalid TMSI. Since the \ac{AMF} cannot link the TMSI to an existing session, it will prompt the UE to provide its SUCI using an \texttt{NAS Identity Request}. The UE replies to that request with its SUCI embedded in the \texttt{NAS Identity Response}, and the attacker can sniff this response being transmitted on the uplink and store both the TMSI and the SUCI that belong to it.

In the later message exchange, the UE is properly authenticated, but the \texttt{NAS Security Mode Command} is rejected by the UE. This is because in the \texttt{NAS Registration Request}, also a set of security capabilities are transmitted by the UE, and if these do not match, the security command procedure will inevitably fail. This is highly beneficial for our attack scenario since the UE will abort the connection establishment and re-try, keeping the TMSI valid. If the registration procedure would instead complete, it would mean that the AMF issues a fresh TMSI to the UE and any connection re-establishment would need to be attacked. This way, the attacker only needs to attack each UE once for as long as the TMSI remains valid. To this end, we set the capabilities in the Registration Request to EA0, EA1, IA1, IA7, and EA7, a combination which is unlikely to be supported by any phone.

An attacker with an access to the operator private keys (e.g., a state actor), could ask for the collected SUCI to be decrypted and obtain the long-term identifiers SUPIs. In that case, the attacker can skip the SUCI linking attack introduced in~\autoref{sec:suci_replay_attack}.

\subsection{SUCI Replay Attack}\label{sec:suci_replay_attack}
This final attack has two distinct use-cases. First, it can be used to link a previously captured SUCI to a UE re-connecting to the cell. Second, it can also be used to cause a DoS or downgrade attack for all UEs, except the UE with a known SUCI, essentially whitelisting the attacker's device.

\begin{figure}[h]
	\centering
	\includegraphics[width=\linewidth]{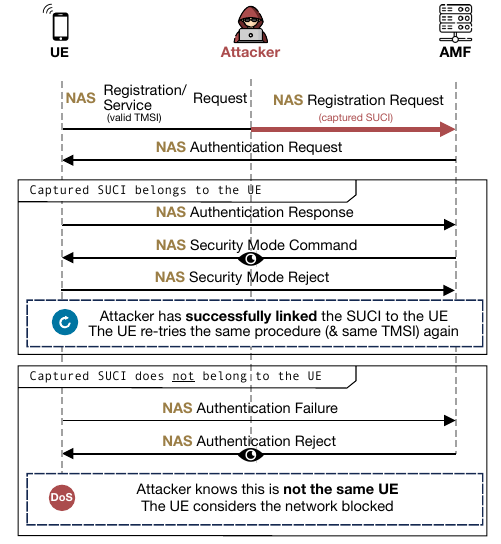}
	\caption{SUCI Replay Attack - Allows the attacker to re-identify a UE belonging to a previously captured SUCI, or causes a persistent DoS at the UE.}
	\label{fig:suci_phase_2}
\end{figure}

To test if a captured SUCI belongs to a UE, the attacker needs to acquire fresh authentication vectors from the AMF for that SUCI, and transmit them to the UE. Then, the attacker can observe from the reaction of the UE (i.e., if the authentication succeeds, or not), if the SUCI indeed belongs to the UE, or not. However, to acquire such fresh authentication vectors previously required an active MITM, using both a fake base station and a UE connecting to the real network and asking for fresh authentication vectors, as described in~\cite{chlosta_5g_2021}. 

However, we can achieve the same result as follows. First, the attacker overshadow the \texttt{NAS Registration / Service Request} with a \texttt{NAS Registration Request} containing the SUCI we have captured in~\autoref{sec:suci_extraction}. The network will react to that with an \texttt{NAS Authentication Request} since the SUCI is valid and belongs to a subscriber, and there is nothing preventing an old SUCI to be re-played, as the authentication vectors are generated independently without binding it to the SUCI, or using any freshness provided by the SUCI. If the SUCI belongs to the UE, the UE will accept the authentication vector and reply with a \texttt{NAS Authentication Response}, resulting in a \texttt{NAS Security Command}, which the attacker can observe and deduce that, indeed, this is the UE belonging to this SUCI. On the other hand, if the SUCI belongs to a different UE, the UE will send an \texttt{NAS Authenticaion Failure}, resulting in an \texttt{NAS Authentication Reject} which again is observable by the attacker, allowing it to deduce that the SUCI does not belong to this UE.

As a secondary effect, the \texttt{Authentication Reject} also causes a downgrade of the UEs not belonging to the replayed SUCI to 4G. On one of the UEs, we have observed that after receiving such a reject, the device does not connect to either 5G or other older cellular standards.

%% file: content/04_implementation.tex
\section{Implementation}

None of the existing open source 5G-SA UE stacks is performant enough to show uplink overshadowing attacks from \autoref{sec:attacks} on the operators base stations (see \autoref{sec:related_work} for more details). In this section, we show the architecture of \name that allowed us to run these attacks reliably against the most latency aggressive gNodeB configurations in the wild. Our whole stacks run on a widely available commodity hardware without need for accelerators or GPU processing.

\subsection{\name Hardware}

\begin{figure}
	\centering
	\includegraphics[width=\linewidth]{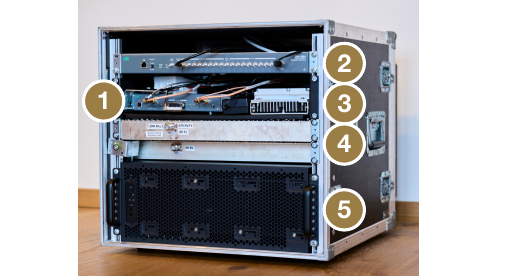}
	\caption{The portable rack, outfitted with: (1) The X310 SDR, the (2) Octoclock GSPDO, (3) B78 and B8 Power Amplifiers, (4) Filters and Combiners, and (5) the AMD 7950X server}
	\label{fig:hw_box}
\end{figure}

In~\autoref{fig:hw_box}, we show our assembled setup in a portable 19-inch rack. Central to the system is the USRP X310 \ac{SDR}. We connected the X310 to the x86 host using the PCIe connectivity kit. The host itself is a COTS machine with an AMD Ryzen 7950X CPU and 64 GB of RAM. The SDR runs on the TDD band n78 (3.4 - 3.8GHz), with a sampling rate of 184.32 MSPS, which is necessary to attack cells with 100 MHz bandwidth. Due to the bandwidth limitations of the PCIe interface, only one Rx and one Tx port can be used at once at this sampling rate. The Rx and Tx are connected to an RF frontend module made by Boostel, which provides an Rx chain with an LNA and a Tx chain with a PA. Since we also support attacks on TDD, only one of the LNA or PA can be enabled at the same time. Thus, whenever there is a scheduled uplink transmission, we switch on one of the GPIO pins on the X310. The GPIO pin is then connected to the PA/LNA enable pin of the Boostel frontend module. The RF frontend is connected to an external band filter. Missing from the picture is the antenna, which would be connected to the front of the filters. A directional multiband 5G antenna (Panorama WMM8G-7-38) is used in the experiments. Finally, To achieve stable frequency and time synchronization, an Octoclock-G provides a 10 MHz reference clock to the X310.

\subsection{Software Architecture}
\begin{figure}[h]
	\centering
	\includegraphics[width=\linewidth]{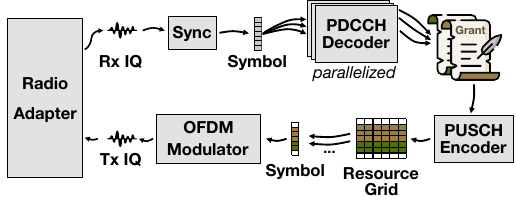}
	\caption{Main components of our symbol-based software architecture.}
	\label{fig:sw_architecture}
\end{figure}
\autoref{fig:sw_architecture} shows the main components of our software architecture. First, the radio adapter handles the transfer of the samples to and from the radio. Second, the cell synchronization time-aligns the samples to the cell, performs an FFT, and forwards single OFDM symbols to the decoder component. Next, if any uplink grants are found, they are encoded on the resource grid and one-by-one modulated and forwarded to the radio adapter.

The main reason why \name is able to achieve sub-500$\mu$s reaction time is the symbol-based processing architecture. That means almost all software components operate on decoding, forwarding, or generating approximately $36\mu s$ worth of samples. Previously, cellular stacks implemented a subframe-processing architecture that operates on chunks of 1 ms, an order of magnitude larger than in \name. 

The architectural decisions explained in this section are the result of continuous performance analysis and optimization. One practical insight is that it is advantageous to have a lot of control over the entire stack, which allowed us to continuously adapt and improve the software architecture to reach the necessary latency and performance goals. Tools to analyze bottlenecks are invaluable, both on the method level, where we developed a large set of benchmarks using Google Benchmark~\cite{google_googlebenchmark_2024}, as well as on the system level, observing the runtime behavior by collecting traces using tracy~\cite{taudul_wolfpldtracy_2026}.

\subsubsection{Radio Adapter Layer}
The radio driver would like to transfer and receive a fixed-sized chunk of samples in regularly spaced intervals. On the other hand, our attacking architecture requires a stream of sample chunks corresponding to the OFDM symbols, while sporadically producing samples in the case of an attack transmission. This attack transmission might be scheduled a few milliseconds in the future, or at other times, merely hundreds of microseconds ahead.

\begin{figure}[h]
	\centering
	\includegraphics[width=\linewidth]{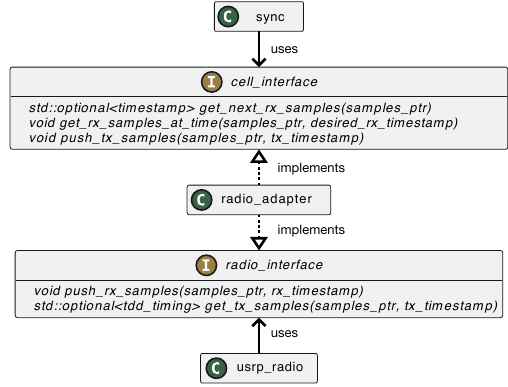}
	\caption{UML class diagram of the Radio Adapter Layer}
	\label{fig:radio_adapter}
\end{figure}

For this reason, the radio adapter layer is introduced, which resolves the mismatch in chunk sizes and the pacing of the received/transmitted samples. As shown in~\autoref{fig:radio_adapter}, it provides an interface for both the radio and the cell, where the radio can submit Rx samples for an arbitrary time and request Tx samples for an arbitrary time. Likewise, for the cell processing object (simply called a cell instance), it can request an arbitrary amount of samples in a continuous fashion (\texttt{get\_next\_tx\_samples}), or for a specific Rx timestamp (\texttt{get\_rx\_samples\_at\_time}), and transmit arbitrarily long, possibly overlapping transmissions at arbitrary timestamps in the future via calls to \texttt{push\_tx\_samples}. Note that many cell instances may exist in parallel.

Crucially, this architecture enables us to independently tune the specific radio driver to achieve the lowest possible Rx-to-Tx delay, while also allowing the cell to operate on the symbol level. For example, with this system, we can, after decoding a DCI message containing an uplink grant in the first symbol of a slot, already start to generate the uplink PUSCH symbols for the grant without having to consider whether the radio might be ready to receive them.

\subsubsection{Cell Sync}

A key requirement for decoding the control traffic and transmitting attack symbols at the correct time is synchronization with a base station. We thus developed a cell synchronization component, which takes as input a stream of IQ samples and outputs a stream of OFDM symbols in the frequency domain. It has two states: the find state, where it finds the synchronization signals, decodes the initial parameters, and synchronizes to the current symbol; and the track state, where it maintains the sync with small adjustments and outputs the 5G symbols.

The initial sync is found by performing a circular correlation using PSS and SSS signals that are part of the SSB. Once the correlation peak exceeds a threshold, we try to decode the MIB that is encapsulated inside the SSB. If the decoding is successful, we have found a sync. Using the MIB, we determine the current symbol and its timestamp. The cell synchronization component then switches to the tracking state.

From the first step, we know which symbol index we received at what timestamp. We can thus determine the timestamp of individual symbols and request from the radio adapter chunks corresponding to those symbols. We strip the received symbols of the cyclic prefix and perform an FFT. After the FFT, we have the values within the individual subcarriers, and we can forward them to the next layers for processing. Every time a symbol contains a PSS or SSS, we check if the correlation peak has drifted. If we measure a drift, we correct the sync by requesting a few extra or fewer samples from the radio adapter, maintaining precise time-alignment.

\subsection{Downlink Decoding}
From the synchronization component, we obtain a single symbol as the output, which is an array of complex float values of the individual subcarriers. These symbols could contain either the \ac{PDCCH} or \ac{PDSCH}, and we need to decode them accordingly. However, while the underlying physical channels are the same, there is a difference in how these channels are used for transporting a \ac{SIB} message, \ac{RAR} message, or even UE-specific messages. Thus, in \name, we implemented distinct components for each of these cases, each separately handling the details of these channels and specifying how to interpret the messages on the channels and how to react, e.g., spawning a new UE instance for a RAR message or encoding an uplink message.

\subsubsection{PDCCH Decoding}
Every component has different needs for the PDCCH decoding; namely, it may need to monitor different search spaces belonging to  different CORESETs, each with different scheduling, scrambling IDs, and various other physical layer parameters. Fundamentally, however, the PDCCH decoding is performed in the following way. First, the necessary symbols from the coreset are aggregated, depending on which symbol the search space starts in and for how many symbols the CORESET is scheduled. Next, it computes which of the subcarriers could potentially carry DCI messages. Only for these subcarriers do we perform channel estimation, equalization, and demodulation. Every potential DCI location is first analyzed by checking the average power of the resulting symbols. Only if the average power is above a certain empirically determined threshold will the location be processed further for potential DCIs. While previous steps are our own implementation, the final steps are implemented using functions from srsRAN~\cite{puschmann_srsltesrslte_2020}. In these final steps, we are descrambling the LLRs and executing the polar decoding, CRC calculation, and validation. Afterwards, the DCI bits are available.

After acquiring the bits, we turn the raw DCIs into downlink or uplink \textit{grants}. We handle various DCI formats for RARs, SIs, and UEs, and for both uplink and downlink allocations. The resulting grant determines which frequency and time resources contain PUSCH or PDSCH. At this point, we can use the decoded grant to decode a PDSCH message sent by the gNodeB using core functions of srsRAN, decode the PUSCH transmission sent by a victim UE, or encode and transmit adversarial PUSCH. In ~\autoref{sec:attacks}, we explain the flow of each attack, i.e., at what point the victim's uplink grant is overshadowed by an adversarial payload, and at what point we decode the victim's transmission instead.

\subsection{Uplink Decoding}

For the SUCI Catching attacks explained in \autoref{sec:suci_extraction}, we need to decode the PUSCH sent by the victim UE. We develop another software component, the uplink decoder, based on the software design of~\cite{kotuliak2025findingphonesfastlowlatency}. Once we decode the uplink allocation grant and the UE has transmitted its PUSCH message, the component fetches relevant IQ samples for the symbols from the radio adapter. In this case, we use the get\_rx\_samples\_at\_time function of the radio adapter since we know the precise time when the transmission happened.

Unlike \cite{kotuliak2025findingphonesfastlowlatency}, we do not extract PHY level characteristics; instead, we decode the uplink messages. Thus, after the FFT of every symbol in the PUSCH, we perform channel estimation of the relevant subcarriers, equalization, and demodulation, followed by decoding them. Finally, if the CRC check passes, we obtain individual bits in the PUSCH transmission. In the SUCI Extraction attack, we are only interested in unencrypted control plane messages; thus, only messages with lcid=1 are further parsed. We reassemble these messages first on the RLC level based on their sequence numbers. Finally, once the whole message for the given RLC sequence number arrives, we can extract the RRC and NAS layer messages. In the SUCI Extraction attack, we are interested only in the Identity Response message. Other NAS-layer messages are ignored.

\subsection{Uplink Transmission}

\subsubsection{DCCCH Message Encoding}
First, the attacker must generate a valid higher-layer message. In all of our NAS-layer attacks in \autoref{sec:attacks}, this is the \texttt{RRC Setup Complete} message, taking into account the RRC Transaction ID from the \texttt{RRC Setup}. Next, the attacker must select the \ac{PLMN} ID under attack from the list of available PLMNs broadcasted on the cell in the SIB1 message. Finally, the attacker must encode the NAS message to inject and embed it in the \texttt{RRC Setup Complete} message.

To transmit this higher-layer message, the attacker must first pass it through the \ac{PDCP} layer. The PDCP layer appends both a sequence number and a message authentication code. Since we overshadow the very first PDCP message, we set the sequence number to 0 and similarly set the 4 bytes of the message authentication code to all zeros as well, since the integrity of the \ac{PDCP} layer has not yet been activated by an \texttt{RRC Security Mode Command} message.

Next, the \ac{PDCP} message must pass through the \ac{RLC} layer. The \ac{RLC} layer does not provide any authentication or encryption; rather, it offers a reliable communication channel capable of transmitting payloads of arbitrary length on top of the lower layers. To this end, the \ac{RLC} layer can fragment the payload into multiple segments, each of which is individually transmitted and acknowledged. In our attacks, we take care not to fragment the payload and only transmit a single segment with the sequence number 0.

\name only transmits a single segment to overshadow a single \ac{NAS} message and still allows subsequent communication on any layer between the UE and the \ac{gNB}. This means that \name takes care not to de-synchronise the RLC layer between the UE and the \ac{gNB}. This means that the attacker needs to \textit{wait} until it receives an uplink grant with sufficient bytes to transmit the encoded RLC/PDCP/RRC/NAS payload in its entirety. If the attacker receives a grant with \textit{fewer} bytes available, it sends a MAC frame containing a buffer status report element, informing the \ac{gNB} that the \ac{UE} has a \textit{lot} of data to transmit imminently. This way, the \ac{gNB} will eventually issue an uplink grant of sufficient size, allowing the attacker to transmit the payload.

This process is repeated until the attacker has managed to transmit the payload at least once. In practice, more re-transmissions will improve the reliability of the attack. After that, the UE transmission may resume. Since the UE might have split the message into more segments than the attacker, the UE might re-transmit these messages. However, since at this point the full message with the same sequence number has already arrived and been processed by the upper layers, in practice, the \ac{gNB} will silently drop the remaining RLC fragments and continue with the message exchange as desired by the attacker. %

\subsubsection{PUSCH Encoding}
To encode the PUSCH, we assume that we have decoded the PDCCH and generated a set of uplink grants, each with a corresponding payload attached to it based on the attack logic. First, we aggregate all grants that are for the same uplink transmission slot together and execute the following process. Starting with the slot nearest in the future, for every grant in the slot, we encode the payload into a PUSCH transmission on the correct subcarriers and symbol locations using srsRAN in the common resource grid.%
The PUSCH encoding process itself is a rather cheap operation since the uplink grants for the first messages in a connection are generally not large and do not span many resource blocks, in comparison with the IFFT operation, which is costly since it always spans the entire resource grid. Thus, this aggregation allows us to execute the costly IFFT and cyclic prefix insertion only once per transmitted symbol of each slot, while still allowing us to attack many UEs in parallel. Immediately after the IFFT and cyclic prefix insertion are done for a symbol, the time-domain symbol is forwarded to the radio adapter, such that it may be sent out while the other time domain symbols are still being generated, achieving the sub-500$\mu$s latency. 

\subsubsection{Timing Advance}
For the uplink overshadowing to work, it is necessary for the attacker to transmit the message precisely at the right time and on the right frequency. Apart from the correct symbol in which the message should be transmitted, each UE also applies its own \textit{timing advance} to each transmission, such that, no matter how far away the UEs are, the base station receives all transmissions at the exact same time. The attacker cannot simply use the victim's timing advance, since the attacker's distance to the \ac{gNB} might be different to the UE's distance from the \ac{gNB}. Applying the wrong timing advance would result in the attacker's message arriving too early or too late. Thus, the attacker also has to determine the timing advance independently. This can be determined either by observing the timing advance values used by a regular UE co-located with the attacker or by measuring the geographical distance from the attacker to the \ac{gNB} and compensating for that.

%% file: content/06_evaluation.tex
\section{Evaluation}
In this section, we evaluate the attacks end-to-end in both lab and real-world settings with a range of COTS 5G UEs as victims. The victim UEs had chipsets from Qualcomm (Samsung S23, OnePlus Pro 10, Nothing Phone (3), iPhone 16 Pro, iPhone 17 Pro), as well as Exynos (Pixel 10 Pro) and Mediatek (Xiaomi 15T Pro).

\subsection{Lab Experiments}
Some of the attacks are not evaluated against a real-world network since we cannot limit their impact to just our own devices. Thus, we re-create the same setting in the lab with an Amarisoft gNodeB ~\cite{amarisoft_amari_2026-1} on B78, running two 100 MHz cells ($\text{gNB}_1\text{, }\text{gNB}_2$, with the power of $\text{gNB}_1$ being 10 dB higher than $\text{gNB}_2$), as well as one LTE cell (eNB) on B7. %

\begin{table}
    \centering
    \begin{tabularx}{\linewidth}{@{} X X X @{}} 
    \toprule
    \textbf{Phone Model} & \textbf{Cell-Wide DoS} & \textbf{Reg. Reject} \\ \midrule
    Samsung S23 & $\text{gNB}_1\rightarrow\text{gNB}_2$ &  $\text{gNB}_1\rightarrow\text{eNB}$\\ 
    OnePlus Pro 10 & $\text{gNB}_1\rightarrow\text{gNB}_2$ &  $\text{gNB}_1\rightarrow\text{eNB}$\\ 
    Nothing Phone (3) & $\text{gNB}_1\rightarrow\text{gNB}_2$ &  $\text{gNB}_1\rightarrow\text{eNB}$\\ 
    iPhone 16 Pro & $\text{gNB}_1\rightarrow\text{gNB}_2$ &  $\text{gNB}_1\rightarrow\text{eNB}$\\ 
    iPhone 17 Pro & $\text{gNB}_1\rightarrow\text{gNB}_2$&  $\text{gNB}_1\rightarrow\text{eNB}$\\ 
    Xiaomi 15T Pro & $\text{gNB}_1\rightarrow\text{gNB}_2$&  $\text{gNB}_1\rightarrow\text{eNB}$\\ 
    Pixel 10 Pro & $\text{gNB}_1\rightarrow\text{DoS}\rightarrow\text{eNB}$ &  $\text{gNB}_1\rightarrow\text{eNB}$\\ 

    \bottomrule
    \end{tabularx}
    \caption{Tests of our phones in our lab environment for the Cell-Wide DoS attack and the Registration Reject Attack.}
    \label{tab:lab_experiments}
\end{table}

\subsubsection{Cell-Wide Denial-of-Service Attack}\label{sec:eval_mac_dos}
This attack cannot be limited to attacking only our own devices since, at the stage when the attack would be executed, we have no identifiers that could be used to bind the connection attempt to our phone. Thus, we executed it under the lab conditions set forth above. We launched the attack on $\text{gNB}_1$ and observed whether the attack worked and what was the behavior of the phones.

The results are summarized in~\autoref{tab:lab_experiments}. None of the phones managed to connect to the cell, despite transmitting between 20 and 100 PRACH preambles before switching to the weaker gNB or giving up. After failing to connect, all phones connected to the weaker gNodeB, except for the Pixel 10 Pro, which completely shut off all cellular services until flight mode was toggled, at which point it re-connected to the LTE cell. We also repeated the experiment in the operator's lab against a commercial gNodeB, which was also operating on B78 with a 100 MHz bandwidth. There, we tested the Samsung S23 SM-S911B/DS, OnePlus Pro 10, and iPhone 16 Pro, resulting in no connection being possible with this gNodeB.

Finally, we evaluated the scalability of the attack. To this end, we connected the $\text{gNB}_1$ with the Amarisoft UE simulator~\cite{amarisoft_amari_2026}. For every PRACH, the gNodeB replied with a \texttt{RAR} message containing an uplink grant with a $k_2 = 3$, leaving the UE to respond exactly \SI{1}{\milli\second} after the last symbol of the PDSCH containing the grant was received. Using the Amarisoft UE simulator, we simulated 64 UEs, all attempting to connect at the same time every 2 seconds over 10 minutes.

\begin{table}
    \centering
    \begin{tabular}{lrrr}
    \toprule
     \textbf{Operation} &       \textbf{Mean} & \textbf{StdDev} &  \textbf{Count}\\
\midrule
    Sync \& Demod & \SI{5.98}{\micro\second} & \SI{1.34}{\micro\second} & \num{2.34E+07}\\ \midrule
    PDCCH (Common) & \SI{7.97}{\micro\second} & \SI{1.10}{\micro\second} & \num{1.67E+06}\\
    PDSCH Decoding & \SI{34.83}{\micro\second} & \SI{4.45}{\micro\second} & \num{1.69E+04}\\ \midrule
    PUSCH Encoding & \SI{6.90}{\micro\second} & \SI{0.99}{\micro\second} & \num{7.35E+04}\\
    Per-Symbol OFDM & \SI{11.97}{\micro\second} & \SI{1.44}{\micro\second} & \num{2.36E+05}\\ \midrule
    \textbf{E2E DL-UL} & \SI{271.72}{\micro\second} & \SI{22.62}{\micro\second} & \num{1.69E+04}\\
    \textbf{Symbol 0 advance} & \SI{672.15}{\micro\second} & \SI{24.34}{\micro\second} & \num{2.36E+05}\\
         \bottomrule
    \end{tabular}
    \caption{Statistics of each processing step while executing a RAR DoS attack against 64 UEs connecting at the same time evvery 2 seconds over 10 minutes.}
    \label{tab:rar_dos_64_ues}
\end{table}

In total, the UEs sent 251976 \texttt{PRACH} messages, a lot of which were overlapping, resulting in 74059 \texttt{RAR} messages, all of them overshadowed by \name successfully. None of the connection attempts succeeded, and all of them were aborted by the gNodeB. During the attack, all relevant function calls were recorded, and their duration collected using tracy~\cite{taudul_wolfpldtracy_2026} and analyzed. Shown in~\autoref{tab:rar_dos_64_ues} is an overview of the relevant operations. First, each symbol is time aligned, and a FFT of size 6144 is performed for each symbol. Next, the common PDCCH search space is decoded, looking for a \texttt{RAR} message on the PDSCH. If one is found, it is decoded, and the grant is extracted. Note that there were, on average, $4.3$ RAR grants contained in each such PDSCH message. Next, each RAR grant is passed to the PUSCH encoder, which encodes the grants and stores them in the frequency domain in the resource grid. Once all grants are encoded, the IFFT and CP insertion are done per-symbol and transmitted to the radio. Most relevant is the metric \textit{Symbol 0 advance} since this tells us how many microseconds ahead the time-domain samples were submitted to the radio layer.
Note that for this attack, only one CPU core was assigned to the cell-level processing, and in this entire chain, no parallelization needed to be implemented.

\subsubsection{Registration Reject Downgrade Attack}

Since we cannot entirely limit the registration reject attack to our own TMSIs, we also executed this attack in our lab. However, first, we explored the behavior of the real-world AMF using the following experiment. Using the app \textit{Network Signal Guru} on a rooted OnePlus Pro 10, we tested a SIM card that had expired to trigger a Registration Reject. We found that the AMF replies with a cause value \#15 (no suitable cells in tracking area). We then configured our lab \ac{gNB} to similarly respond with a \texttt{Registration Reject} with cause value \#15 for subscribers that are not in its database. For each of the phone models tested, we first verified that they were connected to the B78 5G cell and then started the uplink overshadowing attack, triggering the \texttt{Registration Reject \#15}. All of the phones tested switched to 4G within a few seconds, at most 1 minute and 47 seconds, after receiving the \texttt{Registration Reject \#15}, showcasing that the attack produces the intended effect on the UE.

Similarly, we also evaluated the scalability of this attack in the lab against a large number of UEs connecting in parallel. Again, we simulated 64 UEs using the Amarisoft UE simulator, each attempting to establish a connection every 11 seconds for 10 minutes. We configured the gNodeB with search spaces of significant size, spanning all frequency resources of the CORESET and with a total of 14 location candidates across aggregation levels 2 to 16. Note that in this scenario, we configured the gNodeB with $k_2 = 4$, as this is the default supported $k_2$ of the Amarisoft UE.

\begin{table}
    \centering
    \begin{tabular}{lrrr}
    \toprule
     \textbf{Operation} &       \textbf{Mean} & \textbf{StdDev} &  \textbf{Count}\\
\midrule
Sync \& Demod & \SI{5.32}{\micro\second} & \SI{1.26}{\micro\second} & \num{9.34E+06}\\ \midrule
PDCCH (Common) & \SI{8.54}{\micro\second} & \SI{2.60}{\micro\second} & \num{1.11E+07}\\
PDCCH (UE) & \SI{78.72}{\micro\second} & \SI{11.41}{\micro\second} & \num{1.77E+05}\\ \midrule
PDSCH Decoding & \SI{36.32}{\micro\second} & \SI{16.70}{\micro\second} & \num{1.23E+04}\\ \midrule
PUSCH Encoding & \SI{21.78}{\micro\second} & \SI{2.56}{\micro\second} & \num{2.66E+03}\\
Per-Symbol OFDM & \SI{11.04}{\micro\second} & \SI{1.22}{\micro\second} & \num{2.88E+04}\\ \midrule
\textbf{E2E DL-UL} & \SI{368.29}{\micro\second} & \SI{90.04}{\micro\second} & \num{2.06E+03}\\
\textbf{Symbol 0 advance} & \SI{1524}{\micro\second} & \SI{124}{\micro\second} & \num{2.06E+03}\\
         \bottomrule
    \end{tabular}
    \caption{Statistics of each processing step while executing a NAS Registration Reject attack executed against 64 UEs constantly attempting to connect every 11 seconds for 10 minutes.}
    \label{tab:registration_attack}
\end{table}

The results are summarized in~\autoref{tab:registration_attack}. Overall, we can see that the system needs \SI{368.29}{\micro\second} to fully encode and transmit the entire PUSCH in the time-domain, with the first symbol being submitted to the radio layer \SI{1524}{\micro\second} ahead of time. However, for a NAS layer attack to work, the UE-specific PDCCH search space also needs to be decoded. Since this search space is very large with many candidates, it requires a significant amount of time to decode, i.e., \SI{78.72}{\micro\second} on average. For this reason, \name implements the decoding of these UE-specific search spaces in a parallelized manner, utilizing an additional 25 CPU cores scheduled in a round-robin fashion. Only with this optimization is it possible to attack 64 UEs in parallel, a significantly higher portion than what is necessary with a MAC layer attack.

Finally, to evaluate the claim that we can indeed support a $k_2 =1$, we are once again using the Amarisoft as gNodeB, but this time, we configure it with the lowest latency setting possible, i.e., $k_2 = 1$, scheduling an uplink transmission to begin in less than \SI{500}{\micro\second} after the corresponding grant was transmitted on the downlink in the PDCCH. Since the Amarisoft UE does not support these low latencies, we use a real COTS UE, the OnePlus Pro 10, and are using ADB to toggle its flight mode on and off every 0.6 seconds for 13.4 minutes. In total, it attempted to connect 1151 times, with \name failing to attack the connection only once.

\begin{table}
    \centering
    \begin{tabular}{lrrr}
    \toprule
     \textbf{Operation} &       \textbf{Mean} & \textbf{StdDev} &  \textbf{Count}\\
\midrule
Sync \& Demod & \SI{5.13}{\micro\second} & \SI{1.18}{\micro\second} & \num{2.25E+07}\\ \midrule
PDCCH (Common) & \SI{8.27}{\micro\second} & \SI{2.29}{\micro\second} & \num{1.72E+06}\\
PDCCH (UE) & \SI{77.65}{\micro\second} & \SI{10.14}{\micro\second} & \num{7.41E+04}\\ \midrule
PDSCH Decoding & \SI{52.64}{\micro\second} & \SI{22.29}{\micro\second} & \num{4.61E+03}\\ \midrule
PUSCH Encoding & \SI{25.42}{\micro\second} & \SI{4.36}{\micro\second} & \num{2.19E+03}\\
Per-Symbol OFDM & \SI{11.66}{\micro\second} & \SI{1.54}{\micro\second} & \num{3.07E+04}\\ \midrule
\textbf{E2E DL-UL} & \SI{362.96}{\micro\second} & \SI{29.28}{\micro\second} & \num{2.19E+03}\\
\textbf{Symbol 0 advance} & \SI{63.17}{\micro\second} & \SI{26.12}{\micro\second} & \num{2.99E+04}\\

         \bottomrule
    \end{tabular}
    \caption{Statistics of each processing step while executing a NAS Registration Reject 1151 times in a row against OnePlus Pro 10, with $k_2=1$, RTT \SI{200}{\micro\second}, chunk size \SI{25}{\micro\second} during 13.4 minutes. 1 attack failed.}
    \label{tab:nas_registration_performance_k2_1}
\end{table}

The processing times are collected in~\autoref{tab:nas_registration_performance_k2_1}. We can see that in this configuration, the E2E processing time to decode the entire chain from waiting to receive a PDCCH symbol until the last sample of the PUSCH is transmitted to the radio is on average \SI{362.96}{\micro\second}. However, because the radio interface operates with a fixed Rx-Tx delay of \SI{200}{\micro\second}, it is necessary to submit the symbols to the radio layer as soon as the IFFT and CP insertion are completed, which is done on average \SI{63.17}{\micro\second} in advance for the first symbol of the PUSCH transmission.

These results confirm the need for a symbol-based processing architecture and show that \name is capable of attacking UEs even in challenging low-latency PUSCH scheduling scenarios that have previously been out of reach for any software UE implementation. 

\subsection{Real-World Experiments}
For all real-world experiments, we first acquired permission from the operator to execute tests on their production network on our own devices, and we developed the software such that information from other connections on this base station is neither recorded nor stored anywhere. Furthermore, we executed only the attacks on their production network, which we can limit to our phones using the current TMSI. Since we know the TMSI of our own phones, either with the help of the operator or by using the \textit{Network Signal Guru} app, we can limit the attacks entirely to our phones. Therefore, during these tests, neither the privacy nor the availability of any other user was impacted. We also had the experimental plan reviewed by our \ac{IRB}, which confirmed that the experiments could be conducted. During the experiments, we placed the attacker in a favorable position such that the downlink of the chosen cell on B78 with 100 MHz was decodable without issues. The cell has only a limited coverage area with an output power of less than 6 Watts. The results of the experiments are shown in~\autoref{tab:real_world_experiments}.

\begin{table}
    \centering
    \begin{tabularx}{\linewidth}{@{} Y N N N @{}} 
    \toprule
    \textbf{Phone Model} & \textbf{SUCI Catch} & \textbf{SUCI Replay (own)} & \textbf{SUCI Replay (other)} \\ \midrule
    Samsung S23 &   \checkmark    &   \checkmark    &   5G$\rightarrow$4G    \\ 
    OnePlus Pro 10 &  \checkmark    &   \checkmark    &   DoS    \\ 
    Nothing Phone (3) &   \checkmark    &   \checkmark    &  5G$\rightarrow$4G \\ %
    iPhone 16 Pro &   \checkmark    &   \checkmark    &   5G$\rightarrow$4G    \\ 
    iPhone 17 Pro &   \checkmark    &   \checkmark    &   5G$\rightarrow$4G    \\ 
    Xiaomi 15T Pro* &   \checkmark    &   \checkmark    &  5G$\rightarrow$4G     \\ 
    Pixel 10 Pro &   \checkmark    &   \checkmark    &   5G$\rightarrow$5G    \\ 

    \bottomrule
    \end{tabularx}
    \caption{Real-World Experiments executed against our own phone models. Note that the Xiaomi 15T Pro refused to connect to the public 5G SA network, thus we executed the same experiments in our lab.}
    \label{tab:real_world_experiments}
\end{table}

\subsubsection{SUCI Extraction Attack}
For each phone under test, we first connected it to the production 5G-SA network of the operator. We then waited until the phone entered idle mode, reconnected, and launched the uplink overshadowing and uplink sniffing attack. After the attack, we verified that we had first successfully attacked the connection by listening on the downlink for the \texttt{NAS Identity Request} message and then proceeded to sniff the uplink and downlink and decode the SUCI transmitted in the \texttt{NAS Identity Response}. Since we are using B78 for the attacks, which is a TDD band, we can, in the same process, first transmit on the uplink, switch to Rx mode, and keep listening to both downlink and uplink to sniff the identity response.

We tested this attack successfully on the real production \ac{gNB} on all our phone models, except the Xiaomi 15T Pro, for which we executed the attack only in lab conditions since it failed to connect to the public 5G SA network despite numerous attempts to do so.

\subsubsection{SUCI Replay Attack}
For this attack, we conducted the following real-world experiment. First, we connected our phone to the operator network and extracted its TMSI, entering it into the running attack software. We then waited until the phone re-connected to the network and ran the SUCI replay attack. We replayed the SUCI belonging to the same SIM card and verified that the procedure yielded a \texttt{Security Mode Command} and that our system recognized the UE as belonging to this SUCI. This attack worked on all tested phones.

We then repeated the experiment, this time replaying a different SUCI from a second SIM card we obtained from the operator, verified that the procedure terminated with an \texttt{Authentication Reject}, and observed the behavior of the phones. For almost all phones, we observed that receiving the \texttt{Authentication Reject} resulted in a downgrade attack from 5G to 4G, with the phones re-connecting to 4G in between 10 seconds and 1 minute. For the OnePlus Pro 10, this attack resulted in a complete denial of all cellular services that lasted more than 17 hours, after which we stopped the experiment.

%% file: content/07_related_work.tex
\section{Related Work}\label{sec:related_work}

\noindent\textbf{LTE/4G Overshadowing.} Downlink overshadowing of cellular protocols was first introduced and performed against LTE/4G networks in \cite{yang_hiding_2019}. This work introduced an attacker capable of performing denial of service (DoS) by overshadowing the broadcast messages. Further refinements and extensions in LTE/4G overshadowing are presented in~\cite{kotuliak_ltrack_nodate,erni_adaptover_2022, erni_glados_2025, tan_breaking_2022}. Those works managed to go beyond denial of service attacks and introduced privacy attacks along with uplink overshadowing techniques (as opposed to downlink overshadowing only).

\noindent\textbf{5G Overshadowing.} The security of 5G was first formally proven in~\cite{basin_formal_2018, hu_systematic_2019}, and first privacy attacks using rogue base stations were prototyped in~\cite{chlosta_5g_2021}. Follow-up works and studies such as~\cite{ludant_sigunder_2021, luo_sni5gect_2025, hamici-aubert_leveraging_2024} specifically explored downlink overshadowing against 5G and presented a wide variety of attacks that achieve denial of service or decrease of quality of service. The work in~\cite{luo_sni5gect_2025} is most similar to ours in terms of attack types, as it also explores privacy attacks, but facilitated by downlink overshadowing. None of these works explores uplink overshadowing. Uplink overshadowing requires significantly less power and offers a much larger range. The stealthiness of this attack technique is also much higher. However, as explained in this paper, it also requires a significantly different architecture of its platform.

\noindent\textbf{5G Software Platforms.} Most related to our presented platform are the \emph{srsRAN Project}, an open source O-RAN 5G CU/DU solution from Software Radio Systems (SRS)~\cite{puschmann_srsltesrslte_2020,srsRAN_Project2026} and \emph{OpenAirInterface} who implements an open 5G wireless software stack~\cite{OpenAirInterface5G2026}.
Both srsRAN and OpenAirInterface have the main focus of devising implementations for the Radio Access Network (RAN) and the core network in 5G that reach a level of maturity where interoperability with 5G-enabled COTS UEs is seamless and indistinguishable from real commercial offerings such as gNBs from Ericsson, Nokia and Huawei~\cite{Borgini2025_Top5GInfrastructure}.
We note that this objective is different from the goal of our work; in the uplink attacks we present, the adversarial interaction with the gNB is limited to the connection establishment phase, whereas the aforementioned projects aim to provide a holistic experience that can offer, e.g., high-throughput connectivity, setting up data bearers, etc.

Although modifications of srsRAN and OpenAirInterface have been used to evaluate attacks (in the RAN), their focus was on attacks related to the core network~\cite{Giambartolomei2022_Pentesting5GCore}, fake base station attacks~\cite{10978813} and detection~\cite{10.5555/3766078.3766354,electronics13173474}, or penetration testing/fuzzing of 5G systems~\cite{RiedelKoepsell2025_5G_Pentest_UE,10978813,10001673,9606317}.
Compared to srsRAN and other partial implementations of the 5G stack (including free5GRAN~\cite{free5GRAN2026}), OpenAirInterface is the only project that features software code that supports UE functionality.

The unmodified OpenAirInterface code features a sample processing pipeline that is slot-based, i.e., the smallest internal unit to process uplink and downlink is a slice of samples obtained by the SDR that makes up one slot as defined in the numerology of 5G. As a result, the minimum delay that OpenAirInterface requires between the reception of an uplink grant (on the downlink) and the transmission of the corresponding uplink message, i.e., $k_2$, is a multiple of the duration of a slot.
At the time of this work, the delay was set to \emph{three} slots~\cite{OAI_impl_defs_top2026}, which renders the code unusable for uplink attacks in the real-world. Most of the commercial 5G SA depolyments that we encountered use $k_2$ values of 1 or 2, where fast processing is required, and thus hardware acceleration and/or symbol-based processing becomes necessary.

We adopted a symbol-based approach that does not necessitate specialty hardware, other than the software-defined radio to capture the samples.
The OpenAirInterface alliance has added hardware acceleration in the form of the Xilinx T1 chip~\cite{9739186,Xilinx2020_5G_Telco_Accelerator_Cards} and the Nvidia Aerial platform~\cite{villa2024openprogrammablemultivendor5g,NVIDIA_AerialResearchCloud_ManualInstall} for offloading of L1/PHY processing. Moreover, there have been a few independent research-driven initiatives to accelerate processing within OpenAirInterface using FPGAs, such as~\cite{11026993,Romani2020_HW_5G_LDPC_FPGA}.
When paired with hardware acceleration, the OpenAirInterface's implementation of the gNB achieves a downlink throughput of up to $300\unit{Mbps}$~\cite{Villa_2025} and end-to-end latencies as low as $10\unit{ms}$ when connected to COTS UEs~\cite{11317218}. To the best of our knowledge, there is currently no openly available implementation of a software UE or any other software-based solution that meets the latency requirements for uplink overshadowing in the real world.

%% file: content/08_countermeasures.tex
\section{Countermeasures}

Despite the very stealthy nature of overshadowing attacks, there is a range of countermeasures that could be implemented, both on the UE and the gNB sides. The simplest approach is to perform statistical anomaly detection by monitoring connection behavior KPIs such as the number of failed connection attempts or \texttt{Registration / Service / Reject} or \texttt{Identity Requests} messages occurring. 

A more powerful defense would be a comparison of attach procedure transcripts observed by UE and gNodeB (similar to TLS handshake transcript checking) that begins with the very first message and is checked at critical stages, such that any modifications to messages are detected. While denial-of-service is always possible in wireless communications, the UEs can also try to re-connect more aggressively on the same or a different gNodeB. In terms of handling apparent failures, UEs should also place greater priority on remaining on 5G cells to mitigate the impact of downgrade attempts. Preventing privacy violations through SUCI extraction would be possible through changes in the authentication procedures, as has been proposed, e.g., in~\cite{sultan_active_2025}.

To study the attack behavior further, we have also released the detailed logs from the attacks, encompassing the entire stack from PHY to NAS, from all components (UE, attacker, gNB, core network) to foster awareness of these attacks and encourage the development of countermeasures against them.

%% file: content/09_conclusion.tex
\section{Conclusion}

In this paper, we showed that 5G-SA networks are susceptible to uplink overshadowing attacks and pose as much of a threat as more detectable fake base stations. We proposed uplink overshadowing attacks that are capable of achieving denial-of-service, downgrading to older technologies, and violating the privacy of users. We evaluate these attacks using our system running on commodity x86 hardware against a variety of COTS phones across different chipset manufacturers, both in operational network and lab conditions and share detailed logs for future research. To demonstrate these attacks under the most challenging configurations, we had to develop a new 5G stack capable of handling 100 MHz bandwidth and sub-500$\mu$s reaction times. To this end, our system architecture can be used as a blueprint for future research in 5G, not just for uplink overshadowing attacks. Finally, we conclude the paper with possible countermeasures against these attacks but also propose how these attacks could be used to fight against fake base stations.

\section*{Acknowledgements}
The authors thank Rashmi Ambale Maheshwarappa for her help in implementing and extensively testing the software.

%% file: content/a0_ethics.tex
\section*{Ethical Considerations}
We describe our ethical analysis by considering the stakeholders during the research and those affected by the publication of results, then considering the potential positive and negative impacts and the mitigations for possible harm. We ultimately make a concluding justification for our belief in the net benefit of the work. The position described here was validated with our institution's ethical review board, while publication was agreed with the legal body overseeing export control in the local jurisdiction. 

\subsubsection*{Stakeholders and Impacts}

Given the near-ubiquituous usage of cellular telephony, the impacts of \name potentially affect a large swath of the population. We principally categorise those affected as follows:

\textbf{Users} ranging from normal individuals with mobile phones to special bodies operating critical infrastructure, who depend on secure and reliable operation of cellular infrastructure. They can expect for their service not to be made unreliable nor their private information leaked during the research. They also should be made aware of risks to their security that might be identified. While our research did not directly involve any human participants nor seek to collect data about any people or devices, some elements took place alongside legitimate use. Users could be negatively affected if cellular service was impeded or have their privacy damaged if identifiable data were captured about them. Post-publication they could potentially become victims of crime, if previously-unknown attacks become enabled by our study, however also have the opportunity to be better informed of extant risks to cellular technology due to our results. 

\textbf{Industry participants} such as network operators, cellular equipment manufacturers or standards maintainers, who seek to maintain fit-for-purpose networks, devices and services. They can expect their systems to not be impeded by the research efforts and to be advised of security problems that are discovered. Only the network operator we worked with faced potential impact during the research, if their services were degraded by attack testing. By their involvement they were able to conduct their own examination of the attacks and potential remedies. Post-publication industry participants, like users, may experience malicious parties using the attacks described herein, but also gain insight into precise enabling steps, indicators of attacks and countermeasures to inform future improvements.

\textbf{Researchers} who are concerned with understanding the security and applications of cellular technology. They can expect valid and robustly-tested results along with technical openness. The research community benefit from information about vulnerabilities and attacks, along with details of how to overcome technical challenges in testing modern cellular systems. 

\subsubsection*{Mitigations}

Potential negative impacts on users and industry participants were mitigated during research by minimising interaction with public networks and validating attacks carefully before operating on real cells. All principal research was conducted in private networks using base station simulators, research devices and shielded environments to avoid co-mingling with any normal usage. Attacks were subsequently validated with a network operator -- by testing in their private lab environment -- and information exchanged openly about the operating principles, possible consequences and reasonable remedies. Subsequently, some attacks were then permitted for testing on real infrastructure. We only selected to evaluate attacks here that could be targeted specifically to research devices and operated on cells with a very limited range that were not criticial for the availability of the entire network. In order to target researcher devices, some temporary identifiers were needed. These were collected out-of-band via the operator, or on the research devices themselves. Subsequently, identifiers observed over-the-air on the cell were compared against these known values. While not much public traffic was expected on the closed cells, nevertheless all identifiers that did not match were immediately discarded. 

Potential negative impacts after publication are mitigated by the release of technical insights into the engineering hurdles we overcame, but not full source code for launching attacks directly. Our work with a network operator and discussion of countermeasures helps equip industry participants to mitigate misuse in future. 

\subsubsection*{Justifications for Research and Publication}

This work seeks to shed greater light on the security of the latest generation of cellular networks as they become more widespread. By documenting the results of our testing, we improve public understanding of the realities of cellular security and the conditions under which it cannot be relied upon. 

For the research community, comprehensive evaluation of 5G-SA has been challenging as technical complexity made experimentation prohibitively costly. By sharing our experience we aim to reduce this barrier. While we do not release openly available source code, we seek to provide detailed engineering explanations for recreating our work. We believe this should suitably inform the research community for continued work on 5G-SA, yet also minimises the risk of casual access, for malicious parties, to powerful attacks on this major infrastructure. 

On balance, we considered that potential harms towards stakeholders were suitably mitigated throughout the research process and subsequent to publication. This view was mirrored by our ethical review and collaboration with a network operator. As such we concluded that there were no blocking concerns preventing the research or publication. 

%% file: content/a1_openscience.tex
\section*{Open Science}

We do not make our code publicly available, as it falls under multiple license restrictions, such as dual-use restriction in our jurisdiction. Releasing the implementation would enable immediate misuse of the techniques described in this paper, making deployment trivial even for low-skilled adversaries. Instead, we focus on releasing all relevant information to foster further research on detection and prevention mechanisms. We published an example trace of all attacks from the UE, gNB, as well as MME, including all layers from PHY to NAS and above, with full contents of their payloads. This enables industry as well as academic researchers to understand the attacks in detail and develop detection techniques and countermeasures. We make these public at \url{https://github.com/5gone/dataset}.

%% file: main_usenix.bbl
\begin{thebibliography}{10}

\bibitem{3gpp_ts_38304_3gpp_2025}
3GPP~TS 38.304.
\newblock {3GPP} {TS} 38.304 {V19}.0.0 - {User} {Equipment} ({UE}) procedures
  in {Idle} mode and in {RRC} {Inactive} state, October 2025.

\bibitem{3gpp_ts_38321_3gpp_2025}
3GPP~TS 38.321.
\newblock {3GPP} {TS} 38.321 {V19}.0.0, October 2025.

\bibitem{3gpp_ts_38331_3gpp_2025}
3GPP~TS 38.331.
\newblock {3GPP} {TS} 38.331 {V19}.0.0, October 2025.

\bibitem{3gpp_ts_24501_3gpp_2025}
{3GPP TS 24.501}.
\newblock {3GPP} {TS} 24.501 {V18}.12.0, October 2025.

\bibitem{10978813}
Sharique Ahmad, Maik Holzhey, Elif Tasdemir, Mehmet~Akif Kurt, and Frank~H.P.
  Fitzek.
\newblock Practical guidelines to assess vulnerabilities in 5g core network and
  open ran.
\newblock In {\em 2025 IEEE Wireless Communications and Networking Conference
  (WCNC)}, pages 1--6, 2025.

\bibitem{amarisoft_amari_2026-1}
Amarisoft.
\newblock {AMARI} {Callbox} {Series}, 2026.

\bibitem{amarisoft_amari_2026}
Amarisoft.
\newblock {AMARI} {UE} {Simbox} {Series}, 2026.

\bibitem{basin_formal_2018}
David Basin, Jannik Dreier, Lucca Hirschi, Saša Radomirovic, Ralf Sasse, and
  Vincent Stettler.
\newblock A {Formal} {Analysis} of {5G} {Authentication}.
\newblock In {\em Proceedings of the 2018 {ACM} {SIGSAC} {Conference} on
  {Computer} and {Communications} {Security}}, {CCS} '18, pages 1383--1396, New
  York, NY, USA, October 2018. Association for Computing Machinery.

\bibitem{11026993}
Abhishek Bhattacharyya, Andrea Fumagalli, and Koteswararao Kondepu.
\newblock Demo: Fpga-accelerated 5g low-phy functions and an integration with
  openairinterface.
\newblock In {\em 2025 IEEE 26th International Symposium on a World of
  Wireless, Mobile and Multimedia Networks (WoWMoM)}, pages 160--162, 2025.

\bibitem{Borgini2025_Top5GInfrastructure}
Julia Borgini.
\newblock Top 5g infrastructure companies to consider in 2025.
\newblock {\em TechTarget SearchNetworking}, Nov 2025.
\newblock accessed 05 Jan 2026.

\bibitem{chlosta_5g_2021}
Merlin Chlosta, David Rupprecht, Christina Pöpper, and Thorsten Holz.
\newblock {5G} {SUCI}-catchers: still catching them all?
\newblock In {\em Proceedings of the 14th {ACM} {Conference} on {Security} and
  {Privacy} in {Wireless} and {Mobile} {Networks}}, {WiSec} '21, pages
  359--364, New York, NY, USA, June 2021. Association for Computing Machinery.

\bibitem{11317218}
Hubert Djuitcheu, Khurshid Alam, Andrew Sergeev, Florian Spinty, Achim
  Autenrieth, Jörg-Peter Elbers, and Jochen Seitz.
\newblock Industrial open ran deployment: A comparative performance evaluation
  of open and commercial gnbs.
\newblock In {\em 2025 IEEE Future Networks World Forum (FNWF)}, pages 1--6,
  2025.

\bibitem{erni_protocol-aware_2020}
Simon Erni.
\newblock Protocol-{Aware} {Reactive} {LTE} {Signal} {Overshadowing} and its
  {Applications} in {DoS} {Attacks}.
\newblock page~68, September 2020.

\bibitem{erni_glados_2025}
Simon Erni, Martin Kotuliak, Richard Baker, Ivan Martinovic, and Srdjan Capkun.
\newblock {GLaDoS}: {Location}-aware {Denial}-of-{Service} of {Cellular}
  {Networks}.
\newblock pages 5307--5325, 2025.

\bibitem{erni_adaptover_2022}
Simon Erni, Martin Kotuliak, Patrick Leu, Marc Roeschlin, and Srdjan Capkun.
\newblock {AdaptOver}: adaptive overshadowing attacks in cellular networks.
\newblock In {\em Proceedings of the 28th {Annual} {International} {Conference}
  on {Mobile} {Computing} {And} {Networking}}, pages 743--755, Sydney NSW
  Australia, October 2022. ACM.

\bibitem{10001673}
Matheus~E. Garbelini, Zewen Shang, Sudipta Chattopadhyay, Sumei Sun, and Ernest
  Kurniawan.
\newblock Towards automated fuzzing of 4g/5g protocol implementations over the
  air.
\newblock In {\em GLOBECOM 2022 - 2022 IEEE Global Communications Conference},
  pages 86--92, 2022.

\bibitem{Giambartolomei2022_Pentesting5GCore}
Filippo Giambartolomei.
\newblock Penetration testing applied to 5g core network.
\newblock Master's thesis, University of Padova, 2022.
\newblock Master’s thesis, Dipartimento di Matematica “Tullio
  Levi-Civita”, accessed 05 Feb 2026.

\bibitem{google_googlebenchmark_2024}
Google.
\newblock google/benchmark, November 2024.
\newblock original-date: 2013-12-12T00:10:48Z.

\bibitem{hamici-aubert_leveraging_2024}
Virgil Hamici-Aubert, Julien Saint-Martin, Renzo~E. Navas, Georgios~Z.
  Papadopoulos, Guillaume Doyen, and Xavier Lagrange.
\newblock Leveraging {Overshadowing} for {Time}-{Delay} {Attacks} in {4G}/{5G}
  {Cellular} {Networks}: {An} {Empirical} {Assessment}.
\newblock In {\em Proceedings of the 19th {International} {Conference} on
  {Availability}, {Reliability} and {Security}}, {ARES} '24, pages 1--10, New
  York, NY, USA, July 2024. Association for Computing Machinery.

\bibitem{hu_sigoverlay_2025}
Shi Hu, Shan Wang, Jian Wang, Quan Peng, Jingyu Tang, and Jingni Chen.
\newblock {SigOverlay}: {A} {Security} {Evaluation} {Model} {Based} on
  {Reference} {Signals}.
\newblock {\em Computers \& Security}, page 104750, November 2025.

\bibitem{hu_systematic_2019}
Xinxin Hu, Caixia Liu, Shuxin Liu, Wei You, Yingle Li, and Yu~Zhao.
\newblock A {Systematic} {Analysis} {Method} for {5G} {Non}-{Access} {Stratum}
  {Signalling} {Security}.
\newblock {\em IEEE Access}, 7:125424--125441, 2019.
\newblock Conference Name: IEEE Access.

\bibitem{free5GRAN2026}
{IMTSDRLab}.
\newblock {free5GRAN: Open-source 5G RAN stack}.
\newblock \url{https://github.com/IMTSDRLab/free5GRAN}, 2026.
\newblock GitHub repository, accessed 06 Jan 2026.

\bibitem{9739186}
Florian Kaltenberger, Hongzhi Wang, and Sakthivel Velumani.
\newblock Performance evaluation of offloading ldpc decoding to an fpga in 5g
  baseband processing.
\newblock In {\em WSA 2021; 25th International ITG Workshop on Smart Antennas},
  pages 1--4, 2021.

\bibitem{kotuliak_ltrack_nodate}
Martin Kotuliak, Simon Erni, Patrick Leu, Marc Röschlin, and Srdjan Capkun.
\newblock {LTRACK}: {Stealthy} {Tracking} of {Mobile} {Phones} in {LTE}.

\bibitem{kotuliak2025findingphonesfastlowlatency}
Martin Kotuliak, Simon Erni, Jakub Polák, Marc Roeschlin, Richard Baker, Ivan
  Martinovic, and Srdjan Čapkun.
\newblock Finding phones fast: Low-latency and scalable monitoring of cellular
  communications in sensitive areas, 2025.

\bibitem{ludant_sigunder_2021}
Norbert Ludant and Guevara Noubir.
\newblock {SigUnder}: a stealthy {5G} low power attack and defenses.
\newblock In {\em Proceedings of the 14th {ACM} {Conference} on {Security} and
  {Privacy} in {Wireless} and {Mobile} {Networks}}, {WiSec} '21, pages
  250--260, New York, NY, USA, June 2021. Association for Computing Machinery.

\bibitem{luo_sni5gect_2025}
Shijie Luo, Matheus Garbelini, Sudipta Chattopadhyay, and Jianying Zhou.
\newblock {SNI5GECT}: {A} {Practical} {Approach} to {Inject} {aNRchy} into {5G}
  {NR}.
\newblock pages 5385--5404, 2025.

\bibitem{10.5555/3766078.3766354}
Kazi~Samin Mubasshir, Imtiaz Karim, and Elisa Bertino.
\newblock Gotta detect 'em all: fake base station and multi-step attack
  detection in cellular networks.
\newblock In {\em Proceedings of the 34th USENIX Conference on Security
  Symposium}, SEC '25, USA, 2025. USENIX Association.

\bibitem{NVIDIA_AerialResearchCloud_ManualInstall}
{NVIDIA Corporation}.
\newblock Aerial research cloud installation guide: Manual install.
\newblock
  \url{https://docs.nvidia.com/aerial/aerial-research-cloud/archive/a1.0/text/installation_guide/manual_install.html},
  2024.
\newblock NVIDIA Aerial Research Cloud documentation, accessed 04 Jan 2026.

\bibitem{OAI_impl_defs_top2026}
{OpenAirInterface Software Alliance}.
\newblock {impl\_defs\_top.h} — openairinterface5g phy definitions.
\newblock
  \url{https://gitlab.eurecom.fr/oai/openairinterface5g/-/blob/develop/openair1/PHY/impl_defs_top.h},
  2026.
\newblock Source code file in the OpenAirInterface5G repository, accessed 05
  Jan 2026.

\bibitem{OpenAirInterface5G2026}
{openairinterface5G contributors}.
\newblock {openairinterface5G: OpenAirInterface 5G Wireless Implementation}.
\newblock \url{https://gitlab.eurecom.fr/oai/openairinterface5g}, 2026.
\newblock GitLab repository, accessed 05 Jan 2026.

\bibitem{9606317}
Srinath Potnuru and Prajwol~Kumar Nakarmi.
\newblock Berserker: Asn.1-based fuzzing of radio resource control protocol for
  4g and 5g.
\newblock In {\em 2021 17th International Conference on Wireless and Mobile
  Computing, Networking and Communications (WiMob)}, pages 295--300, 2021.

\bibitem{electronics13173474}
Sourav Purification, Jinoh Kim, Jonghyun Kim, and Sang-Yoon Chang.
\newblock Fake base station detection and link routing defense.
\newblock {\em Electronics}, 13(17), 2024.

\bibitem{puschmann_srsltesrslte_2020}
Andre Puschmann, Ismael Gomez, Pedro Alvarez, Xavier Arteaga, Francisco
  Paisana, Paul Sutton, and Justin Tallon.
\newblock {srsLTE}/{srsLTE}, April 2020.
\newblock Last accessed: 28.04.2020.

\bibitem{RiedelKoepsell2025_5G_Pentest_UE}
Richard Riedel and Stefan K{\"o}psell.
\newblock 5g-pentest-ue: A penetration testing framework for identifying 5g
  system vulnerabilities.
\newblock In {\em Availability, Reliability and Security. ARES 2025}, volume
  15999 of {\em Lecture Notes in Computer Science}, pages 326--339. Springer,
  Cham, 2025.

\bibitem{Romani2020_HW_5G_LDPC_FPGA}
Luca Romani.
\newblock Hardware acceleration of 5g ldpc using datacenter-class fpgas.
\newblock Master's thesis, Politecnico di Torino, 2020.
\newblock Master’s thesis, Electronic Engineering, supervisors: Luciano
  Lavagno and Salvatore Scarpina; accessed 05 Feb 2026.

\bibitem{srsRAN_Project2026}
{srsran}.
\newblock {srsRAN\_Project: Open source O-RAN 5G CU/DU solution}.
\newblock \url{https://github.com/srsran/srsran_project}, 2026.
\newblock GitHub repository, accessed 05 Jan 2026.

\bibitem{sultan_active_2025}
Nazatul~H. Sultan, Xinlong Guan, Josef Pieprzyk, Wei Ni, Sharif Abuadbba, and
  Hajime Suzuki.
\newblock Active {Attack} {Resilience} in {5G}: {A} {New} {Take} on
  {Authentication} and {Key} {Agreement}, July 2025.

\bibitem{tan_breaking_2022}
Zhaowei Tan, Boyan Ding, Jinghao Zhao, Yunqi Guo, and Songwu Lu.
\newblock Breaking {Cellular} {IoT} with {Forged} {Data}-{Plane} {Signaling}:
  {Attacks} and {Countermeasure}.
\newblock {\em ACM Transactions on Sensor Networks}, April 2022.
\newblock Just Accepted.

\bibitem{taudul_wolfpldtracy_2026}
Bartosz Taudul.
\newblock wolfpld/tracy, January 2026.
\newblock original-date: 2020-03-17T19:43:04Z.

\bibitem{villa2024openprogrammablemultivendor5g}
Davide Villa, Imran Khan, Florian Kaltenberger, Nicholas Hedberg, Ruben~Soares
  da~Silva, Anupa Kelkar, Chris Dick, Stefano Basagni, Josep~M. Jornet, Tommaso
  Melodia, Michele Polese, and Dimitrios Koutsonikolas.
\newblock An open, programmable, multi-vendor 5g o-ran testbed with nvidia arc
  and openairinterface, 2024.

\bibitem{Villa_2025}
Davide Villa, Imran Khan, Florian Kaltenberger, Nicholas Hedberg, Rúben~Soares
  da~Silva, Stefano Maxenti, Leonardo Bonati, Anupa Kelkar, Chris Dick, Eduardo
  Baena, Josep~M. Jornet, Tommaso Melodia, Michele Polese, and Dimitrios
  Koutsonikolas.
\newblock X5g: An open, programmable, multi-vendor, end-to-end, private 5g
  o-ran testbed with nvidia arc and openairinterface.
\newblock {\em IEEE Transactions on Mobile Computing}, 24(11):11305–11322,
  November 2025.

\bibitem{EO13010_1996}
{William J. Clinton}.
\newblock Executive order 13010: Critical infrastructure protection.
\newblock Presidential Executive Order, July 1996.
\newblock Published as \textit{Critical Infrastructure Protection}, 61 Fed.
  Reg. 37347 (July 17, 1996).

\bibitem{Xilinx2020_5G_Telco_Accelerator_Cards}
{Xilinx, Inc.}
\newblock {Xilinx 5G Telco Accelerator Cards}.
\newblock Technical presentation / white paper, Xilinx, Inc., 2020.
\newblock accessed 05 Jan 2026.

\bibitem{yang_hiding_2019}
Hojoon Yang, Sangwook Bae, Mincheol Son, Hongil Kim, Song~Min Kim, and Yongdae
  Kim.
\newblock Hiding in {Plain} {Signal}: {Physical} {Signal} {Overshadowing}
  {Attack} on {LTE}.
\newblock In {\em 28th {USENIX} {Security} {Symposium} ({USENIX} {Security}
  19)}, pages 55--72, Santa Clara, CA, August 2019. USENIX Association.

\end{thebibliography}
